\documentclass[11pt]{article}
\usepackage{moriond,epsfig}

\def\be{\begin{equation}}
\def\ee{\end{equation}}
\def\bea{\begin{eqnarray}}
\def\eea{\end{eqnarray}}

\begin{document}
\vspace*{4cm}
\title{TESTING TIME REVERSAL INVARIANCE WITH COLD NEUTRONS}

\author{T. SOLDNER}

\address{Institut Laue Langevin, BP 156, F-38042 Grenoble Cedex 9, France}

\maketitle\abstracts{
The neutron is a well-suited system to search for a violation of time
reversal invariance beyond the Standard Model. Recent experiments and
projects searching for time reversal violation in the neutron decay
and in the neutron electric dipole moment will be presented.
}

\section{Introduction}

CP violation is implemented in the Standard Model (SM) of particle
physics via a non-trivial phase in the Cabbibo-Kobayashi-Maskava
matrix \cite{kobayashi1973}. Dedicated experiments and facilities
have been built to check the predictions of the SM for CP
violation, e.g. in the Kaon or the B meson sectors. Up to now, no
significant deviations have been found. However, the CP violation implemented
in the SM is insufficient to explain the baryon-antibaryon
asymmetry observed in the Universe \cite{dine2004}. Therefore, new
sources of CP violation are searched for in various systems.

The neutron is well-suited to search for CP violation beyond the SM.
The influence of the SM CP violation is so tiny that it will hardly
ever be observed, whereas several scenarios beyond the SM permit
observable effects. These effects can be searched for in properties
of the neutron, namely in its electric dipole moment (EDM), or in
triple correlations between the decay products. CP or T (time reversal)
violation would become manifest in deviations of these values from 0. 
Therefore these effects would have an unambiguous experimental
signature.

\section{T violating observables}

The EDM of an elementary particle describes the
shift between the centre of the charge distribution and the centre
of mass, i.e.
\begin{equation}
  {\bf d}=\int {\rm d}^3r \rho({\bf r})({\bf r} - {\bf r}_0),
\end{equation}
where $\rho({\bf r})$ denotes the charge distribution and ${\bf r}_0$ the
centre of mass of the particle. For particles with spin ${\bf \sigma}$,
the EDM must be directed parallel or antiparallel to the spin for
symmetry reasons. The quantity ${\bf \sigma}{\bf d}$ therefore is a
P pseudo-scalar and a T pseudo-scalar. If T invariance holds, a particle
can not have spin and EDM at the same time.

The EDMs of atoms and elementary particles measured up to now are
compatible with 0. Particularly the EDM of the electron and the neutron
give some of the most stringent limits for T violation beyond the SM.
For the neutron, the EDM is predicted to be in the range of
$10^{-32}\ldots10^{-31}\,e\cdot{\rm cm}$ in the SM \cite{khriplovich1982}.
The present experimental limit is
$|{\bf d}_n|< 0.63\cdot10^{-25}\,e\cdot{\rm cm}$ (90\% c.l.) \cite{pdg2002}.
An overview about predictions from theories beyond the SM can be found
in \cite{edmproposallosalamos2002}.

\begin{figure}
\psfig{figure=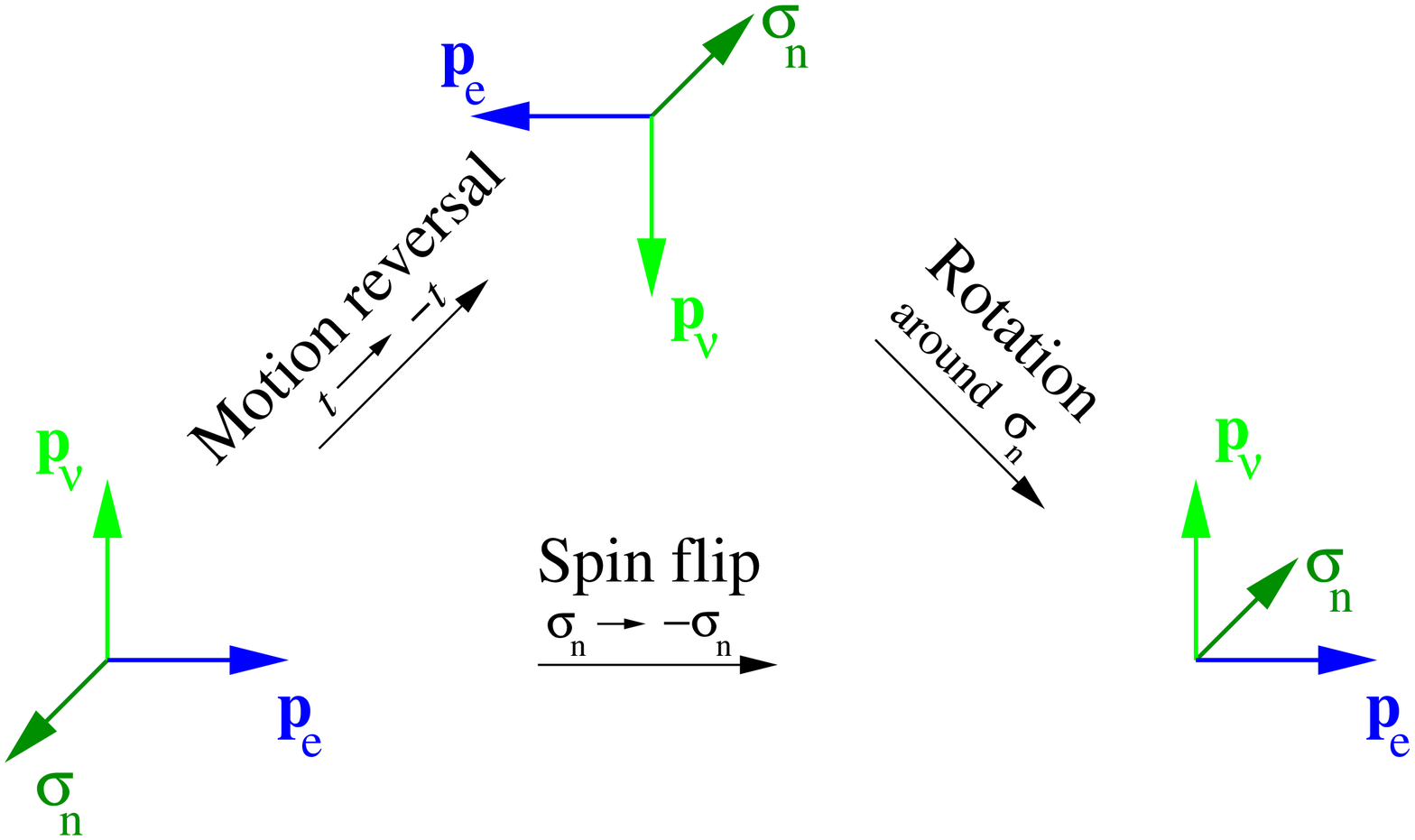,height=4cm}\hfill
\psfig{figure=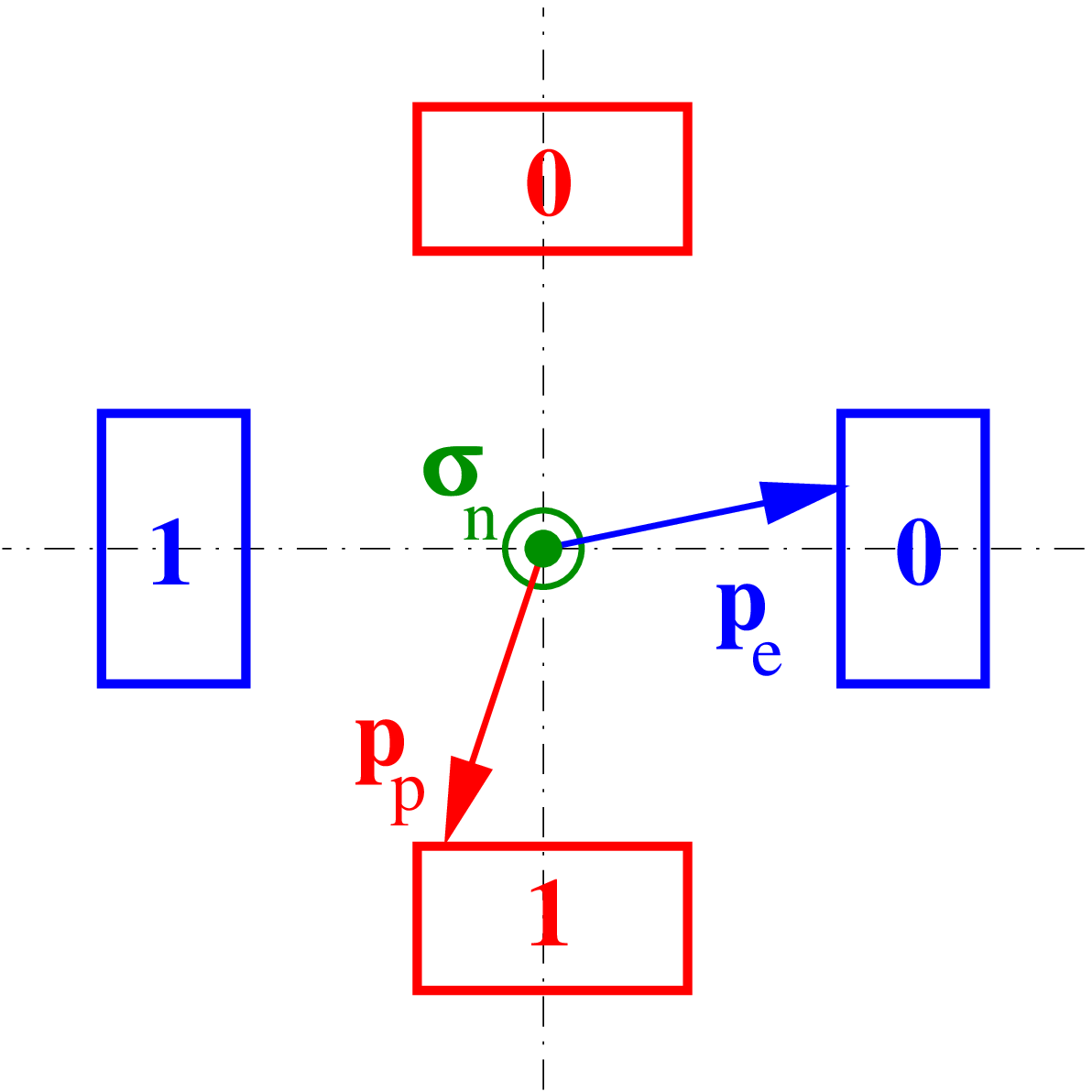,height=4cm}
\caption{Motion reversal for the $D$ correlation (left) and basic set-up
  for a $D$ measurement.\label{fig:motionrev}}
\end{figure}

The neutron decays with a half live of 885.7(8)\,s \cite{pdg2002}
into a proton, an electron, and an electron antineutrino. In this decay,
different triple correlations between the decay products allow to
search for time reversal violation \cite{jackson1957a}, e.g. the correlations
${\bf \sigma}_n({\bf p}_e\times{\bf p}_\nu)$ between the spin of the decaying
neutron and the momenta of electron and antineutrino ($D$ correlation)
or ${\bf \sigma}_e({\bf p}_e\times{\bf \sigma}_n)$ between the spin and the
momentum of the electron and the spin of the decaying neutron
($R$ correlation). These correlations violate motion reversal.
This is demonstrated in Fig.~\ref{fig:motionrev}
for the $D$ correlation: The operation $t\rightarrow-t$ inverts all vectors
(which are first derivatives in time). The resulting diagram can not
be transfered to the initial one by continuous transformations. On the
other hand, motion reversal can be simulated by flipping the neutron spin.

Time reversal is motion reversal and the exchange of initial and final state.
Since the latter can not be realised in this decay, final state effects have
to be taken into account. They can mimic a finite correlation which is not
due to time reversal violation. These effects were calculated 
to \cite{bytev2002} $D_{\rm fs}\approx10^{-5}$ and
\cite{jackson1957b,vogel1983} $R_{\rm fs}\approx 10^{-3}$ \footnote{The
numerical values refer to neutron decay. The same correlations can be
(and have been) investigated in nuclear beta decay. Here, the final state
effects are different and in general larger than for the neutron.}.
Experimental results above the final state effects would be unambiguous
proofs for T violation.
The world average for $D$ given in \cite{pdg2002} is $D=-0.6(1.0)\cdot10^{-3}$.
$R$ has not yet been measured in neutron decay. The prediction 
\cite{herczeg1997} from the SM is $D,R\approx10^{-12}$.

\section{Physics sensitivity}

For the neutron EDM as well as for the correlations $D$ and $R$, the
SM values are many orders of magnitude below the present experimental
precision. Theories beyond the SM, however, allow effects up to the
present experimental limits. A recent detailed comparison of the
sensitivity of $D$ and $R$ for physics beyond the SM with that of EDMs of
atoms and the neutron can be found in \cite{herczeg2001}. In general,
the EDM is the more sensitive parameter, but the $D$ correlation
puts more stringent limits on leptoquark models.
Tab.~\ref{tab:sensitivities} lists values allowed by different scenarios.
The $R$ correlation
(at the presently attempted experimental precision)
can be more sensitive than EDMs only for fine-tuned parameters in some
theories \cite{herczegtalknist2004}.
\begin{table}
  \centerline{\begin{tabular}{lr@{}lr@{}l}\hline
    Model\rule{0pt}{12pt} & &$D$&
      &$d_{\rm n}~[e\cdot{\rm cm}]$\\[1pt]\hline
    Phase in CKM-Matrix\rule{0pt}{12pt}
      & $\approx$&$10^{-12}$ &&$10^{-33}\ldots10^{-31}$\\
    QCD $\theta$ Parameter &  &&&$3\cdot10^{-16}\theta_{\rm QCD}$\\
    SuSy & $\le$&$10^{-7}\ldots10^{-6}$ (from  \cite{lising2000})
      & &$\le$present limits\\
    Left-right symmetric  & $\le$&$10^{-5}\ldots10^{-4}$ &&$\le$present limits
      ($^{199}$Hg)\\
    Exotic fermions        & $\le$&$10^{-5}\ldots10^{-4}$&&$\le$present limits
      ($^{199}$Hg)\\
    Leptoquark                & $\le$&present limits &\\[1pt]\hline
    Experiment \cite{pdg2002}\rule{0pt}{12pt}        &
      --&0.6(1.0)$\cdot10^{-3}$
      &$<$&$0.63\cdot10^{-25}$\\[1pt]\hline
  \end{tabular}}
  \caption[]{Sensitivity of $D$ and $d_{\rm n}$ for new physics. Values from
    \cite{herczeg2001,edmproposallosalamos2002}. For left-right symmetric
    models and exotic fermions, the limit from the EDM of $^{199}$Hg is
    more stringent than that of the neutron EDM.\label{tab:sensitivities}}
\end{table}

It should be noted that $D$ is parity conserving whereas $R$ is parity
violating. Therefore $D$ is sensitive to vector- and axial-vector-type,
$R$ to scalar- and tensor-type T violating interactions. These interactions
can appear on the tree level in $D$ and $R$ but only on the loop level
in EDMs. Therefore, the limits from the correlations in the decay
are theoretically cleaner than the limits from the EDM \cite{herczeg2001}.

\section{Experiments}

\subsection{$D$ correlation}

The measurement principle for the $D$ correlation can be understood from
Fig.~\ref{fig:motionrev}: The coincidence count rates electron and proton
(which is detected instead of the antineutrino) have to be measured
for both directions of the neutron spin. The experiment is complicated
by two reasons: On the one hand, the decay particles have low energies and
are difficult to detect. Especially the proton with a maximal kinetic energy
of 750 eV has to be accelerated by an electrostatic potential to
energies of about 30 keV prior detection. On the other hand, other
correlations in the decay can create systematic effects.

The full differential decay
probability for polarised neutrons and the observation of electron and
antineutrino momenta is \cite{jackson1957a}:
\begin{equation}\label{eqn:JacksonFormel}
  \frac{{\rm d}W}{{\rm d}E_{\rm e}{\rm d}\Omega_{\rm e} 
    {\rm d}\Omega_{\bar\nu}} = 
    g G_{\rm E}(E_{\rm e})
    \left\{1+a\frac{{\bf p}_{\rm e}{\bf p}_{\bar\nu}}{E_{\rm e}E_{\bar\nu}} +
    b\frac{m_{\rm e}}{E_{\rm e}}+{\bf P}\left( 
    A\frac{{\bf p}_{\rm e}}{E_{\rm e}} +
    B\frac{{\bf p}_{\bar\nu}}{E_{\bar\nu}} +
    D\frac{{\bf p}_{\rm e}\times{\bf p}_{\bar\nu}}{E_{\rm e}E_{\bar\nu}}\right)
    \right\}.
\end{equation}
$\bf P$ is the polarisation of the neutrons, $E_i$, ${\bf p}_i$, and $\Omega_i$
the energy, momentum, and solid angle of electron e and antineutrino $\bar\nu$,
$g$ a normalisation constant, $G_{\rm E}$ the electron spectrum, and
$m_{\rm e}$ the electron mass. The correlation coefficients $a$, $b$, $A$,
$B$, and $D$ are related to the coupling constants of the operators describing
nuclear beta decay (see \cite{jackson1957a}) and have to be determined
experimentally. Since $A\approx-0.1$ and 
$B\approx1$ are parity violating, these correlations are the main sources
for systematic errors in a $D$ measurement and have to be suppressed
carefully in the experiment.
Therefore, $D$ measurements work with a symmetric set-up of electron and
proton detector as shown in Fig.~\ref{fig:motionrev}. The neutron beam
is perpendicular to the plane of the drawing and longitudinally polarised.
The set-up has two perpendicular mirror planes of the detectors and the
decay volume (dash-dotted lines). 
The correlation ${\bf P}({\bf p}_{\rm e}\times{\bf p}_{\bar\nu})$
changes its sign under reflection on these planes whereas
${\bf P}{\bf p}_{\rm e}$ and ${\bf P}{\bf p}_{\bar\nu}$ remain unchanged.
Therefore, the combination
\begin{equation}
  \alpha_D:=\alpha^{00}-\alpha^{01}-\alpha^{10}+\alpha^{11} = 
    4D{\bf P}{\bf \kappa}^{00}_D
  \qquad{\rm with}\qquad
  \alpha^{ij}=
    \frac{N_\uparrow^{ij}-N_\downarrow^{ij}}{N_\uparrow^{ij}+N_\downarrow^{ij}}
\end{equation}
of the asymmetries $\alpha^{ij}$ of the four detector combinations
is sensitive to $D$ but not to $A$ and $B$. Deviations from these
symmetries are sources for systematic effects. $N_{\uparrow\downarrow}^{ij}$
denote the count rates for the both spin directions, ${\bf \kappa}_D$
the sensitivity of a detector combination to $D$.

\begin{figure}
\psfig{figure=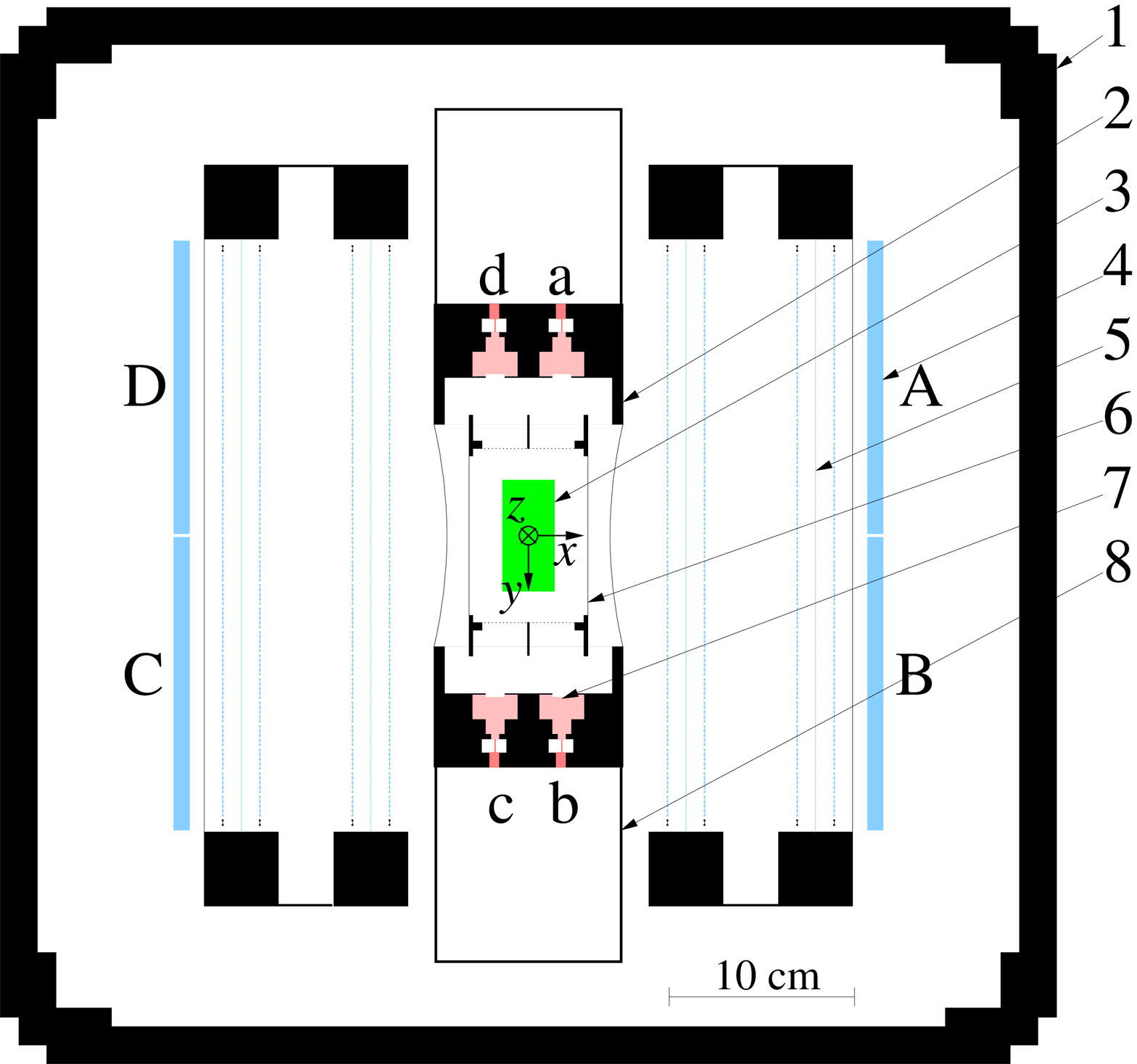,height=5cm}\hfill
\psfig{figure=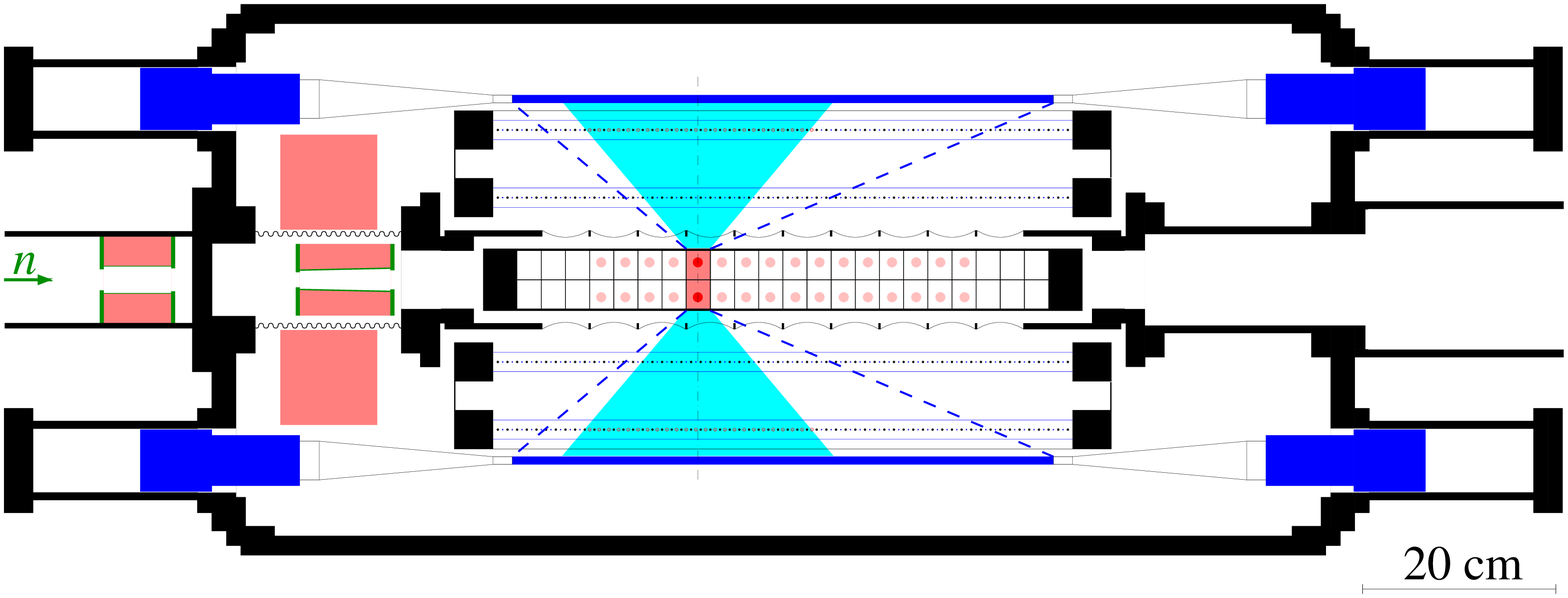,height=3.5cm}
\caption{Left: Cross-section of the Trine detector:
  1 -- detector chamber containing the counting gas, 2 -- inner vacuum chamber,
  3 -- neutron beam, 4 -- plastic scintillator, 5 -- \mbox{MWPC},
  6 -- electrode for proton acceleration, 7 -- PIN diode,
  8 -- housing for PIN preamplifier. The polarisations points in $z$ direction
  perpendicular to the plane of the drawing. Right: Top view.\label{fig:trine}}
\end{figure}

The presently most precise measurement of $D$ was carried out by the
Trine collaboration \cite{soldner2004}. The detector is shown in 
Fig.~\ref{fig:trine}.
The electrons are detected by plastic scintillators in coincidence with
multi wire proportional chambers (MWPCs), the protons after acceleration by
special low noise PiN diodes. The detector consists of 16 planes
defined by 4 PiN diodes per plane which use all the same MWPCs and scintillators
(see top view in Fig.~\ref{fig:trine}). The wire chambers permit an off\/line
selection of a symmetric range for electron detection (shaded area in 
Fig.~\ref{fig:trine}), depending on the
PiN diode hit by the proton. This increase in detector 
symmetry compared to the minimum requirements further
reduces the influence of the parity violating correlations $A$ and $B$.

The final results is \cite{soldner2004}
$D=(-2.8\pm6.4^{\rm stat}\pm3.0^{\rm syst})\cdot10^{-4}$.
The leading systematic errors are due to asymmetries in
electron detection. A measurement with improved statistics and systematics
was carried out in 2003 at the Institut Laue Langevin and is presently
under analysis.

The emiT collaboration uses a detector geometry that is optimised for the
highest statistical sensitivity to $D$.
The kinematics of neutron decay favours large angles between electron and
proton. The statistical sensitivity for $D$ is maximum for angles
of about 135$^\circ$ between electron and proton detectors. The emiT detector
uses an octagonal detector arrangement, resulting in this optimum angle.
The collaboration published a first result
\mbox{$D=(-6\pm12^{\rm stat}\pm5^{\rm syst})\cdot10^{-4}$} in 2000 
\cite{lising2000}. In 2003 a run with improved detector performance was
carried out at the National Institute of Standards and 
Technology \cite{mumm2004}.

\subsection{$R$ correlation}

The $R$ coefficient was not yet measured in neutron decay, but in several
nuclear decays (e.g. \cite{sromicki1996}). The technical difficulty
consists in the detection of the polarisation of the electron. This can
be done by Mott scattering on thin Gold or Lead foils, in combination
with tracking of the electron. An $R$ measurement for neutron decay is
in preparation at the Paul Scherrer Insitut, Villingen. The experiment
attempts for a precision of $10^{-3}$ in the final run.
A first data run is planned for 2004. Details about the apparatus
can be found in \cite{bodek2001}.

\subsection{Neutron EDM}

All recent experiments and proposed projects for the neutron EDM use
the Ramsey resonance method \cite{smith1957}, with ultra-cold
neutrons (UCNs\footnote{UCNs are defined as neutrons with an energy low
enough that they can be stored in material bottles.}). The precision of the 
running experiment is limited by the UCN density available from existing
sources (and by systematic effects). Proposals for new experiments
therefore rely on the significant progress in the UCN density possible
with upcoming super-thermal UCN sources (see e.g.
\cite{edmproposallosalamos2002}).

The statistical sensitivity for the neutron EDM is given by
\begin{equation}
  \sigma_{d_{\rm n}} = \frac{\hbar\sigma}{\alpha E T \sqrt{N}},
\end{equation}
with $\alpha$ describing the efficiency of polarisation and polarisation
analysis, $E$ the electric field applied to the neutron, $T$ the time
the neutrons spend in this field, and $N$ the number of neutrons.
All proposed projects attempt to increase $N$ by orders of magnitude.
Some increase of $E$ is also planned by exploiting liquid $^4$He as
buffer liquid, but this improvement is limited to a few 10 kV/cm.

A completely different approach to measure the neutron EDM
was proposed in \cite{fedorov1997}. In non-centrosymmetric crystals, much
higher electric fields than can be created in vacuum
are provided by nature. For the (110)-plane
of quartz this field has been measured \cite{alexeev1989} to
$E=1.8(2)\cdot10^5$\,kV/cm, about 4 to 5 orders
of magnitude above the fields available in the laboratory. Also the
neutron density for cold neutrons to be used in this experiment is an
order of magnitude above the state-of-the-art UCN density. On the other hand,
the time the neutrons spend in the crystal is short and can be maximised by
using Bragg angles close to 90$^\circ$. In Tab.~\ref{tab:crystaledm}
the sensitivity of this Laue diffraction method is compared with the most
precise UCN EDM experiment.
\begin{table}
  \centerline{\begin{tabular}{lr@{}lr@{}l}\hline
    \rule{0pt}{12pt} & &UCN method \cite{harris2000}&
      &Laue diffraction \cite{fedorov1997,fedorov2004}\\[1pt]\hline
    $\alpha$\rule{0pt}{12pt}    && 0.5 && 0.5\\
    $E$~[kV/cm]                 && 4.5 && $1.8\cdot10^5$\\
    $T$~[s]                     && 130   && $1.3\cdot10^{-3}$\\
    $N$~[neutrons/s]            && 62    && 500\\[1pt]\hline
    ${\rm d}\sigma_{d_n}/{\rm d}t$~$[10^{-25}e\cdot {\rm cm}/{\rm day}]$
            \rule{0pt}{12pt}    &&  6 && 6\\[1pt]\hline
  \end{tabular}}
  \caption[]{Sensitivity of the UCN method (status 2000) and the Laue
    diffraction method for the neutron EDM.\label{tab:crystaledm}}
\end{table}

\begin{figure}
  \centerline{\psfig{file=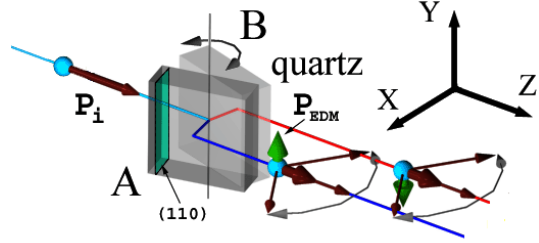,height=3.5cm}}
  \caption{Principle of the Laue diffraction method.\label{fig:lauediffraction}}
\end{figure}
The principle of the method is shown in Fig.~\ref{fig:lauediffraction}.
A longitudinally polarised neutron beam is Laue diffracted by the (110)
plane of a quartz crystal. In this case, the diffracted beam inside the
crystal can be described as superposition of two type of Bloch waves,
one travelling in the plane of the nuclei, the other in the plane
between. Both beams are exposed to opposite electric fields. In their
rest frames, the neutrons are exposed to opposite magnetic fields, originating
from the electric field in the crystal, and precess in contrary directions.
Therefore, the effective polarisation of the diffracted beam
depends on the crystal thickness but stays always in the scattering plane.
A neutron EDM would now lead to the appearance of a polarisation $P_{\rm EDM}$
perpendicular to this plane. This polarisation is related to the neutron
EDM as follows:
\begin{equation}
  P_{\rm EDM} =  \frac{2c^2}{v_\parallel}\frac{d_n}{\mu_n} =
  6\cdot10^{20}\frac{d_n}{e\cdot {\rm cm}},
\end{equation}
with $c$ the speed of light, $\mu_n$ the neutron magnetic moment,
and $v_\parallel$ the component of the neutron velocity parallel to the (110)
plane. The numerical value is given for a thickness of the quartz crystal
of 3.5\,cm, corresponding to a precession of $\pi/2$ for the both Bloch waves,
i.e. a complete depolarisation of the diffracted neutron beam. A first test
of the statistical sensitivity of the method resulted in 
${\rm d}\sigma_{d_{\rm n}}/{\rm d}t = 6\cdot10^{-25}e\cdot{\rm cm}/{\rm day}$
\cite{fedorov2004}. A full scale test of the statistical sensitivity and
systematic effects was carried out recently at the Institut Laue Langevin.

The Laue diffraction method provides a statistical sensitivity which is
comparable with the state-of-the-art UCN EDM experiment but a completely
different set of systematics. Even if it should not be competitive with the
proposed UCN projects, it is worthwhile to exploit this technique for
checks with independent systematics.

\section{Summary}

Low energy searches for physics beyond the SM can compete with
direct searches for new particles. Particularly sensitive are searches for
asymmetries which would not exist in the SM (or have unobservable small values)
but which may be created by contributions from new interactions at orders of
magnitude above the SM values. Examples are the EDM of the neutron and
time reversal violating correlations in neutron decay (especially the $D$
coefficient). Great progress is to expect for the neutron EDM from
different new projects; correlation measurements have been improved recently
and will provide new results within one year.

\section*{References}
\bibliography{moriond}
\end{document}